\documentclass[a4paper,10pt]{article}

\usepackage[margin=2.5cm]{geometry}
\usepackage{graphicx}
\usepackage{times}
\usepackage{url}
\usepackage{hyperref}
\usepackage{booktabs}
\usepackage{enumitem}

\hypersetup{
  colorlinks=true,
  linkcolor=black,
  citecolor=black,
  urlcolor=blue
}

\title{Pedagogical Reflections on the Holistic Cognitive Development (HCD) Framework and AI-Augmented Learning in Creative Computing}

\author{
  Anand Bhojan\\[2pt]
  Department of Computer Science, School of Computing\\
  National University of Singapore\\
  \texttt{banand@comp.nus.edu.sg}
}

\date{2025}

\begin{document}
\maketitle

\begin{abstract}
This paper presents an expanded account of the Holistic Cognitive Development (HCD) framework for reflective and creative learning in computing education. The HCD framework integrates design thinking, experiential learning, and reflective practice into a unified constructivist pedagogy emphasizing autonomy, ownership, and scaffolding. It is applied across courses in game design (CS3247, CS4350), virtual reality (CS4240), and extended reality systems, where students engage in iterative cycles of thinking, creating, criticizing, and reflecting. The paper also examines how AI-augmented systems such as iReflect, ReflexAI, and Knowledge Graph--enhanced Large Language Model (LLM) feedback tools operationalize the HCD framework through scalable, personalized feedback. Empirical findings from previous and ongoing studies demonstrate improved reflective depth, feedback quality, and learner autonomy. The work advocates a balance of supportive autonomy in supervision, where students practice self-directed inquiry while guided through structured reflection and feedback.
\end{abstract}


\section{Introduction}

Designing and developing interactive systems such as games and extended reality (XR) environments requires both technical competence and an empathetic understanding of human experience. Game and XR designers must reason about player psychology, emotion, and meaning-making while also managing complex technical constraints. This makes the learning process inherently experiential, iterative, and multidisciplinary.

My teaching across three core modules at the National University of Singapore---CS3247 (Game Development), CS4350 (Game Development Project), and CS4240 (Virtual Reality Systems)---is guided by the Holistic Cognitive Development (HCD) framework. HCD emphasizes four interrelated learning activities:
\emph{Thinking}, \emph{Creating}, \emph{Criticizing}, and \emph{Reflecting}. These activities are supported by a supervision style of \emph{balanced supportive autonomy} that combines \emph{autonomy}, \emph{ownership}, and \emph{scaffolding}.

In parallel, our group has developed a family of AI-supported tools---including iReflect, ReflexAI, and Knowledge Graph--enhanced LLM feedback engines---to scale feedback and reflection in large classes while preserving depth and quality \cite{bhojan2024ireflect,bhojan2025reflexai,koh2025icer,yu2025cp5105}. These tools operationalize key HCD ideas: they make thinking visible, support critique, and nudge students towards richer reflection.

This paper synthesizes these strands into a unified narrative. We first present the HCD framework and situate it within established theories of design thinking \cite{brown2008}, experiential learning \cite{kolb1984}, and reflective practice \cite{schon1983}. We then describe how HCD shapes the design of creative computing modules and assessments, and how AI-augmented feedback systems instantiate HCD at scale. Finally, we discuss evaluation findings and implications for computing education.

\section{The Holistic Cognitive Development Framework}

\subsection{Learning Activities Cycle}

HCD conceptualizes students' cognitive engagement as a recurring cycle of four learning activities:

\begin{enumerate}[label=\textbf{\arabic*)}]
  \item \textbf{Thinking:} analysing the problem space, articulating goals, and exploring constraints and opportunities.
  \item \textbf{Creating:} generating and implementing design ideas in the form of prototypes, levels, mechanics, or interactive experiences.
  \item \textbf{Criticizing:} engaging in structured critique from peers, teaching staff, and AI tools; interrogating underlying assumptions and trade-offs.
  \item \textbf{Reflecting:} consolidating insights, connecting experience to theory, and planning concrete changes for the next iteration.
\end{enumerate}

Rather than being a linear sequence, these activities form a continuous loop in which reflection feeds back into new thinking and creation. For example, students in CS3247 might begin with an initial design concept (thinking), build a playable prototype (creating), subject it to playtesting and critique (criticizing), and then write a structured reflection on lessons learned (reflecting) before iterating on the design.

\subsection{Supervision Style: Balanced Supportive Autonomy}

The second layer of HCD describes the supervision style required to support this cycle. Students are expected to practice self-directed learning and take responsibility for their decisions, but they also require carefully calibrated guidance. HCD therefore emphasizes \emph{balanced supportive autonomy}, comprising three intertwined dimensions:

\begin{itemize}
  \item \textbf{Autonomy:} students are encouraged to set their own goals, choose themes and mechanics aligned with their interests, and experiment with novel ideas.
  \item \textbf{Ownership:} students are accountable for design decisions, team processes, and the quality of their reflections; they are treated as junior professionals.
  \item \textbf{Scaffolding:} teaching staff provide structure through milestones, rubrics, critique frameworks, and reflective prompts, gradually fading support as competence grows.
\end{itemize}

\begin{figure}[t]
    \centering
    % Replace the filename below with the actual name of your image file on Overleaf.
    \includegraphics[width=0.75\linewidth]{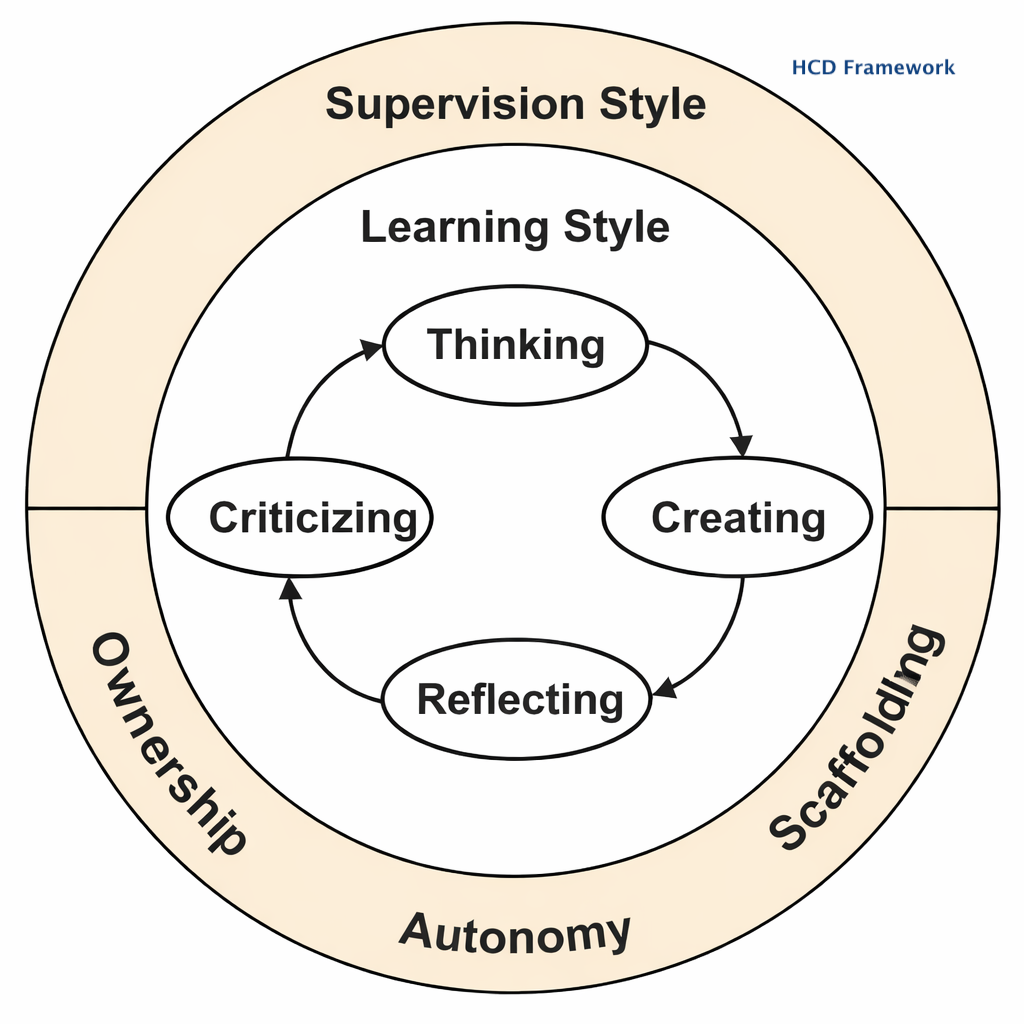}
    \caption{The Holistic Cognitive Development (HCD) framework. The inner learning cycle comprises four iterative activities---Thinking, Creating, Criticizing, and Reflecting. The outer ring represents the supervision style of balanced supportive autonomy, combining Autonomy, Ownership, and Scaffolding. Arrows from the outer ring into the cycle indicate how supervision shapes and supports each learning activity.}
    \label{fig:hcd_framework}
\end{figure}

Figure~\ref{fig:hcd_framework} depicts HCD as two concentric layers: an inner cycle of learning activities surrounded by an outer ring representing supervision.

In practice, balanced supportive autonomy means that supervision is neither laissez-faire nor overly controlling. Instead, it flexes to students' needs: novices receive more explicit modelling and structured feedback, while advanced teams are given greater freedom to define their own evaluation criteria and reflective questions.

\section{Theoretical Foundations}

HCD draws on several complementary traditions in educational theory.

\subsection{Experiential Learning and Reflection}

Kolb's experiential learning cycle \cite{kolb1984} frames learning as a process of moving between concrete experience, reflective observation, abstract conceptualization, and active experimentation. HCD aligns closely with this model: creating corresponds to concrete experience, criticizing and reflecting map to observation and conceptualization, and subsequent iterations constitute experimentation.

Schön's work on the reflective practitioner \cite{schon1983} further sharpens the role of reflection, distinguishing \emph{reflection-in-action} (thinking on one's feet during practice) from \emph{reflection-on-action} (analysis after the fact). In HCD, both forms are cultivated: students are prompted to make design decisions visible during development and then to analyse them afterwards in written reflections.

Earlier scholars such as Dewey emphasized reflective thinking as central to education \cite{dewey1933}. Boud et al.\ \cite{boud1985} and Mezirow \cite{mezirow1991} highlighted reflection as a mechanism for transforming experience into deeper understanding and, eventually, shifts in perspective. HCD positions reflective writing and discussion as vehicles for such transformative learning in computing contexts.

\subsection{Design Thinking and Creative Practice}

Design thinking provides a complementary lens focused on creativity, empathy, and iteration. Brown's influential account \cite{brown2008} articulates stages of inspiration, ideation, and implementation. In HCD, thinking and creating correspond to these phases: students empathise with users, generate ideas, and build artefacts that can be tested and refined.

Crucially, critique and reflection are treated not as add-ons but as central design activities. Structured critique sessions help students to interrogate design assumptions, consider diverse perspectives, and rehearse professional feedback practices. This aligns with work on critical reflection in professional education \cite{rolfe2001}.

\section{Implementation in Creative Computing Courses}

\subsection{Course Contexts}

Table~\ref{tab:courses} summarizes how HCD shapes three core modules.

\begin{table}[t]
  \centering
  \caption{Application of the HCD framework across three creative computing modules.}
  \label{tab:courses}
  \begin{tabular}{lll}
    \toprule
    \textbf{Course} & \textbf{Focus} & \textbf{HCD Application} \\
    \midrule
    CS3247 & Game Development &
      Iterative prototyping, playtesting, and peer critique cycles. \\
    CS4350 & Game Development Project &
      Capstone teams produce a full game, with structured reflections and AI-supported feedback. \\
    CS4240 & Virtual Reality Systems &
      Empathy-centred XR design with emphasis on reflective experience reports. \\
    \bottomrule
  \end{tabular}
\end{table}

Across these modules, assessments are explicitly designed to demand the HCD activities. Typical elements include:

\begin{itemize}
  \item \textbf{Critical reviews} of games or VR experiences, encouraging analytical thinking and articulation of design principles.
  \item \textbf{Playtesting and peer review} sessions, where teams gather feedback from classmates and external players.
  \item \textbf{Reflective journals} and structured reflection prompts that require students to interpret feedback and describe concrete changes for subsequent iterations.
  \item \textbf{Portfolio-style documentation} linking design decisions to theory and playtest evidence.
\end{itemize}

\subsection{Supervision Practices}

Supervision practices are aligned with the HCD supervision ring:

\begin{itemize}
  \item \emph{Autonomy} is supported by allowing teams to choose genres, target audiences, and technologies that match their interests, within broad constraints.
  \item \emph{Ownership} is emphasised through team contracts, milestones, and public showcases (e.g., internal expos), where students present and defend their work.
  \item \emph{Scaffolding} is provided via checklists, design templates, rubrics, and exemplars, as well as stepwise introduction of reflective tools.
\end{itemize}

This combination encourages students to internalise professional standards while preserving the freedom to explore creative directions.

\section{AI-Augmented Reflection}

\subsection{iReflect and Automated Reflective Feedback}

To support reflective practice at scale, we developed the iReflect platform \cite{bhojan2024ireflect,bhojan2024llmreflect}. iReflect is a web application through which students upload playtesting feedback, respond to it, and write reflections. The system was initially designed to centralise peer reviews and reflective writing in creative media courses.

Subsequent work extended iReflect with LLM-based feedback. Using GPT-4 models and reflection rubrics, the system automatically scores student reflections and provides formative comments. Studies in creative media courses have shown that higher-quality reflective writing correlates with stronger project outcomes \cite{bhojan2024ireflect}.

\subsection{ReflexAI: Prompt-Engineered LLM Feedback}

ReflexAI \cite{bhojan2025reflexai} builds on this foundation by systematically optimising prompts to improve LLM feedback quality and consistency. Key techniques include:

\begin{itemize}
  \item \textbf{Few-shot prompting}, where sample reflections and scores are embedded in the prompt to align the model with human raters.
  \item \textbf{Repeated evaluation with averaging}, where the same reflection is scored multiple times at a higher temperature and the mean score is used to reduce randomness.
  \item \textbf{Constructive feedback schemas}, which instruct the model to provide specific strengths, areas for improvement, concrete examples, and forward-looking suggestions.
\end{itemize}

Experiments show that these strategies increase correlation with human raters and produce more actionable feedback compared to naive prompting.

\subsection{Knowledge Graph--Enhanced LLM Feedback}

A further extension augments LLMs with domain knowledge via a lightweight Knowledge Graph (KG) and retrieval-augmented generation framework inspired by LightRAG \cite{guo2024lightrag,yu2025cp5105}. In this setup, information about genres, mechanics, design patterns, and common pitfalls is encoded as a graph. When a student submits playtesting notes or a reflection, relevant nodes and relations are retrieved and added to the LLM prompt \cite{koh2025icer}.

This KG-enhanced approach allows feedback to be more context-aware. For example, a project using physics-based platforming can receive comments linked to prior examples of similar mechanics, known issues with difficulty spikes, or best practices for onboarding. Evaluation results suggest that KG-enhanced feedback is more specific and better aligned with expert comments than generic GPT-4 responses \cite{yu2025cp5105}.

\section{Evaluation and Findings}

The various studies underpinning iReflect, ReflexAI, and KG-enhanced LLM feedback provide converging evidence that AI can effectively support HCD-style learning:

\begin{itemize}
  \item In creative media courses, average reflective writing scores improved significantly after students received AI-generated formative feedback, with effect sizes corresponding to moderate-to-large gains in rubric-based measures \cite{bhojan2024llmreflect,bhojan2025reflexai}.
  \item Correlation between LLM scores and expert ratings reached values around $r = 0.7$--$0.8$ after prompt optimisation and repeated evaluation \cite{bhojan2025reflexai}, approaching the level of agreement commonly reported between human raters.
  \item KG-enhanced feedback reduced variance in LLM scoring and produced comments that students and instructors judged to be more specific and helpful than baseline GPT-4 outputs \cite{yu2025cp5105,koh2025icer}.
  \item Survey data indicate that a majority of students perceived AI feedback as timely, fair, and motivational, and reported greater awareness of their own strengths and weaknesses as designers \cite{bhojan2024ireflect}.
\end{itemize}

From an HCD perspective, these findings suggest that AI tools can effectively expand the scaffolding layer of supervision while preserving autonomy and ownership: students remain responsible for interpreting feedback and deciding how to act on it.

\section{Discussion}

Integrating AI feedback into HCD raises both opportunities and challenges. On the positive side, AI systems make it feasible to provide rich, timely feedback even in large cohorts, allowing students to iterate more frequently. They also model reflective language and critique structures, which students can appropriate and adapt.

However, several risks must be managed:

\begin{itemize}
  \item \textbf{Over-reliance:} students may be tempted to treat AI feedback as authoritative rather than as one perspective among many. HCD supervision practices therefore emphasise triangulation between peer, instructor, and AI feedback.
  \item \textbf{Bias and alignment:} LLM outputs can reflect biases in their training data. Rubrics and prompts must be carefully designed and periodically audited.
  \item \textbf{Authenticity of reflection:} if students attempt to outsource reflection to AI tools, the learning benefits may diminish. Our guidelines explicitly forbid using AI to generate reflections; instead, AI is framed as a tool to support human reflection.
\end{itemize}

Despite these caveats, we argue that AI-augmented feedback is a natural extension of the HCD supervision layer. It allows the scaffolding component to adapt dynamically to students' needs, while human teachers focus on higher-level coaching, pastoral care, and the cultivation of professional identity.

\section{Conclusion}

The Holistic Cognitive Development framework offers a coherent, theory-informed approach to teaching creative computing. By explicitly linking cycles of thinking, creating, criticizing, and reflecting with a supervision style of balanced supportive autonomy, HCD provides both students and educators with a shared language for learning.

The integration of AI-augmented feedback tools such as iReflect, ReflexAI, and Knowledge Graph--enhanced LLM engines demonstrates how HCD can be operationalised at scale. These systems extend the scaffolding available to students while preserving autonomy and ownership. Empirical findings indicate that such tools enhance reflective writing quality, feedback consistency, and metacognitive awareness.

Looking forward, we plan to further investigate how HCD-informed AI systems can adapt to individual learners over time, support cross-course longitudinal reflection, and be applied beyond game and XR design to other areas of computing education.

\end{document}